\documentclass[fleqn,usenatbib]{mnras}
\usepackage{newtxtext,newtxmath}
\usepackage[T1]{fontenc}
\usepackage{anyfontsize}
\DeclareRobustCommand{\VAN}[3]{#2}
\let\VANthebibliography\thebibliography
\def\thebibliography{\DeclareRobustCommand{\VAN}[3]{##3}\VANthebibliography}
\usepackage{graphicx}	
\usepackage{amsmath}	
\title[The outbursts of comet 17P/Holmes]{Long-term outburst activity of comet 17P/Holmes and constraints on ejecta size distributions}
\author[Maria Gritsevich et al.]{
Maria Gritsevich$^{1,2,3}$\thanks{E-mail: maria.gritsevich@helsinki.fi},
Marcin Weso\l{}owski$^{4}$\thanks{E-mail: mWesołowski@ur.edu.pl}, Josep M. Trigo-Rodríguez$^{5,6}$\thanks{E-mail: trigo@ice.csic.es}, Alberto J. Castro-Tirado$^{2,7}$\thanks{E-mail: ajct@iaa.es},
\newauthor{Jorma Ryske$^{8}$, Markku Nissinen$^{8}$, Peter Carson$^{9}$}\\
$^{1}$Faculty of Science, University of Helsinki, Gustaf Hallströmin katu 2, FI-00014 Helsinki, Finland\\
$^{2}$Instituto de Astrofísica de Andalucía (IAA-CSIC), Glorieta de la Astronomía s/n, E-18008, Granada, Spain\\
$^{3}$Institute of Physics and Technology, Ural Federal University, Mira str. 19, 620002 Ekaterinburg\\
$^{4}$University of Rzesz\'ow, Faculty of Exact and Technical Sciences, Institute of Physics, Pigonia 1 Street, 35-310 Rzesz\'ow, Poland\\
$^{5}$Institut de Ciències de l’Espai (ICE, CSIC), Campus UAB, Carrer de 
Can Magrans S/N, Cerdanyola del Vallès, 08193, Catalonia, Spain\\
$^{6}$Institut d’Estudis Espacials de Catalunya (IEEC), Esteve Terradas 1, 
Edifici RDIT, Of. 212, Parc Mediterrani de la Tecnologia (PMT), \\ Campus 
del Baix Llobregat – UPC, Castelldefels (Barcelona), 08860, Catalonia, 
Spain\\
$^{7}$Ingeniería de Sistemas y Autom\'atica, Universidad de M\'alaga, Unidad Asociada al CSIC por el IAA, Escuela de Ingenier\'ias Industriales,\\ Arquitecto Francisco Pe\~nalosa, 6, Campanillas, 29071 M\'alaga, Spain\\
$^{8}$Ursa Astronomical Association, Kopernikuksentie 1, FI-00130 Helsinki, Finland\\
$^{9}$British Astronomical Association Comet Section, London, United Kingdom\\
}
\date{Accepted 2026 March 11. Received 2026 March 7; in original form 2026 February 5}
\pubyear{\the\year{}}
\begin{document}
\label{firstpage}
\pagerange{\pageref{firstpage}--\pageref{lastpage}}
\maketitle
\begin{abstract}
A quantitative understanding of cometary outbursts requires robust constraints on the size distribution of ejected particles, which governs outburst dynamics and underpins estimates of released gas and dust. In the absence of direct measurements of particle sizes, assumptions about the size distribution play a central role in modelling dust-trail formation, their dynamical evolution and observability, and the potential production of meteor showers following encounters with Earth. We analyse brightness amplitude variations associated with outbursts of comet 17P/Holmes from 1892 to 2021, with particular emphasis on the exceptional 2007 mega-outburst. During this event the comet underwent a rapid and substantial brightening: at its peak, the expanding coma reached a diameter exceeding that of the Sun and briefly became the largest object in the Solar System visible to the naked eye. We constrain the size distribution and total mass of porous agglomerates composed of ice, organics, and dust ejected during the outburst. 
The inferred particle size distribution is consistent with a power law of index $q$, yielding effective particle sizes ranging from $\sim 10^{-6}\,\mathrm{m}$ (for $q = 4$) to $\sim 5 \times 10^{-3}\,\mathrm{m}$ (for $q = 2$). Accounting for effective particle size, sublimation flux, and bulk density, we find that the total number of ejected particles increases with both \( q \) and sublimation flux. These results place constraints on the physical properties of outburst ejecta and provide physically motivated initial conditions for long-term dust-trail evolution modelling. They further indicate that cometary outburst brightness is determined primarily by the number of particles and their size distribution, rather than by the total ejected mass alone, with direct implications for the origin and evolution of meteoroid streams and the interplanetary dust population.

\end{abstract}

\section{Introduction}
\label{sec:1}
 
Cometary outbursts are sudden, energetic events during which comets eject large quantities of gas and dust, producing rapid and sometimes dramatic increases in brightness \citep{Hughes1990,Prialnik1995,Filonenko2006,Miles2016a,Miles2016b,Hajra2017,Skiff2018,Kelley2021,Kelley2022,Belousov2024a,Belousov2024b,Muller2024,Wesołowski2021,Wesołowski2022a,Wesołowski2025}. These events provide unique opportunities to probe the physical and chemical properties of cometary nuclei, the mechanisms driving episodic activity, and the processes responsible for the formation of dust trails and meteoroid populations in the Solar System.
 
The largest outburst ever recorded in observational cometary history is that of comet 17P/Holmes (hereinafter 17P) in October 2007, occurring nearly six months after its perihelion passage \citep{Sekanina2007,Moreno2008,Montalto2008,Stevenson2010,Stevenson2012,Gronkowski2010,Lacerda2012,Wesołowski2020,Wesołowski2021,Wesołowski2022a,Gritsevich2022,Nissinen2025}. Observations documented a rise in brightness from magnitude 16.5 to 2.5 within approximately 42 hours, corresponding to an increase of 14 mag, or a brightening by a factor of $\sim 4 \times 10^{5}$. At its peak, the expanding coma exceeded the diameter of the Sun, despite the total mass of the comet remaining negligible in comparison.

Quantifying the mass ejected during such outbursts is crucial for understanding cometary composition, internal structure, evolutionary pathways, dust trail formation, and the potential contribution of comets to meteoroid populations \citep{Gritsevich2025b}. Since the 2007 event, many studies have attempted to estimate the mass ejected from 17P using a range of observational data and modelling approaches. One of the earliest analyses was conducted by \cite{Sekanina2007}, who interpreted the outburst as the explosive fragmentation of a portion of the nucleus and derived mass estimates based on the expansion of the dust coma. Using photometric observations, \cite{Montalto2008} subsequently estimated the dust production rate and total ejected mass, suggesting values in the range 10$^{10}$--10$^{12}$~kg.

Later studies employed increasingly sophisticated modelling techniques to refine these estimates. \cite{Stevenson2010} incorporated particle size distributions and accounted for the effects of radiation pressure and solar wind on the dust particles, providing a more detailed description of the outburst dynamics. \cite{Wesołowski2020,Wesołowski2022a} further advanced these efforts through numerical simulations that coupled dust dynamics with observational constraints. \cite{Gritsevich2022} developed an advanced modelling framework that simulates both the initial particle release and the long-term evolution of the resulting dust trail, producing validated predictions for particle positions in space for three distinct size populations \citep{Ryske2022}. This approach demonstrates that meteoroid trail particles can be located in space, tracked over multiple revolutions after release, and identified as being observable, enabling their direct association with parent bodies and even with the specific outburst from which they originated. Such methods provide a complementary and cost-effective alternative to dedicated space missions. Building on this foundation, future in situ dust analyzers could further improve our understanding of small Solar System bodies \citep{Kruger2024}, particularly since many observed dust trails may originate from previously undetected or weakly active comets \citep{Sykes1986,Moreno2025}.

Despite these advances, substantial uncertainties remain. Major challenges include accurately constraining the physical properties of the ejected material, characterizing the initial conditions of the outburst, and accounting for the complex interplay of forces acting on particles after ejection  \citep{Nissinen2025}. In particular, the role of sublimation processes, especially water ice sublimation, in regulating the mass loss and subsequent fragmentation of ejected material remains insufficiently constrained. These agglomerates, consisting of dust initially embedded within volatile phases, are expected to progressively disintegrate into micron-sized particles upon solar exposure, as proposed for comet 29P/Schwassmann–Wachmann \citep{Trigo2008b,Trigo2010,Wesołowski2025a}. The disruption of such aggregates into fine particles leads to a significant increase in the cumulative optical cross-section, accompanied by corresponding photometric brightening. This effect may provide a means to anticipate the onset of an outburst, which may initially manifest as enhanced thermal emission from large aggregates, producing a measurable increase in the $I$-band brightness.

Simplifications adopted in previous models may therefore introduce substantial uncertainties in the derived mass estimates. These challenges are further compounded by the limited availability of high-cadence, high-resolution observations during the most intense phases of activity. For instance, the maximum rate of brightening in the 2007 outburst occurred approximately 1.2 days after the onset of activity \citep{Li2011}, further complicating the interpretation of the early evolution of the ejecta.

In this study, we develop a model that incorporates the key physical and chemical processes governing cometary outbursts and the subsequent mass ejection. We present selected results from our observational campaigns and compile and analyze data from all documented outbursts of 17P, including observations from its most recent return. Using the observed brightness amplitudes from 1892 to 2021, with particular emphasis on the extraordinary 2007 event, we constrain the size distribution and total mass of porous dust agglomerates ejected from 17P. Our results provide improved constraints on the physical properties of outburst ejecta and establish a physically motivated basis for modeling dust trails, as well as for assessing the broader role of episodic cometary activity in shaping meteoroid populations in the Solar System.

\section{Observations}

17P was first observed during a bright outburst on 1892 November 6 by Edwin A. Holmes from suburban London while conducting routine observations of the Andromeda Galaxy \citep{Holmes1892}. Independent discoveries followed shortly thereafter, with Thomas David Anderson observing the comet on November 8 from Edinburgh, Scotland, and John Ewen Davidson on November 9 from Mackay, Queensland, Australia. A second major outburst, first reported by Johann Palisa, occurred in mid-January 1893. During both events, the comet appeared as a luminous, rapidly expanding disk visible to the naked eye, gradually becoming more diffuse before fading into the background sky over subsequent weeks.

\begin{figure}
\begin{center}
\includegraphics[width=9cm]{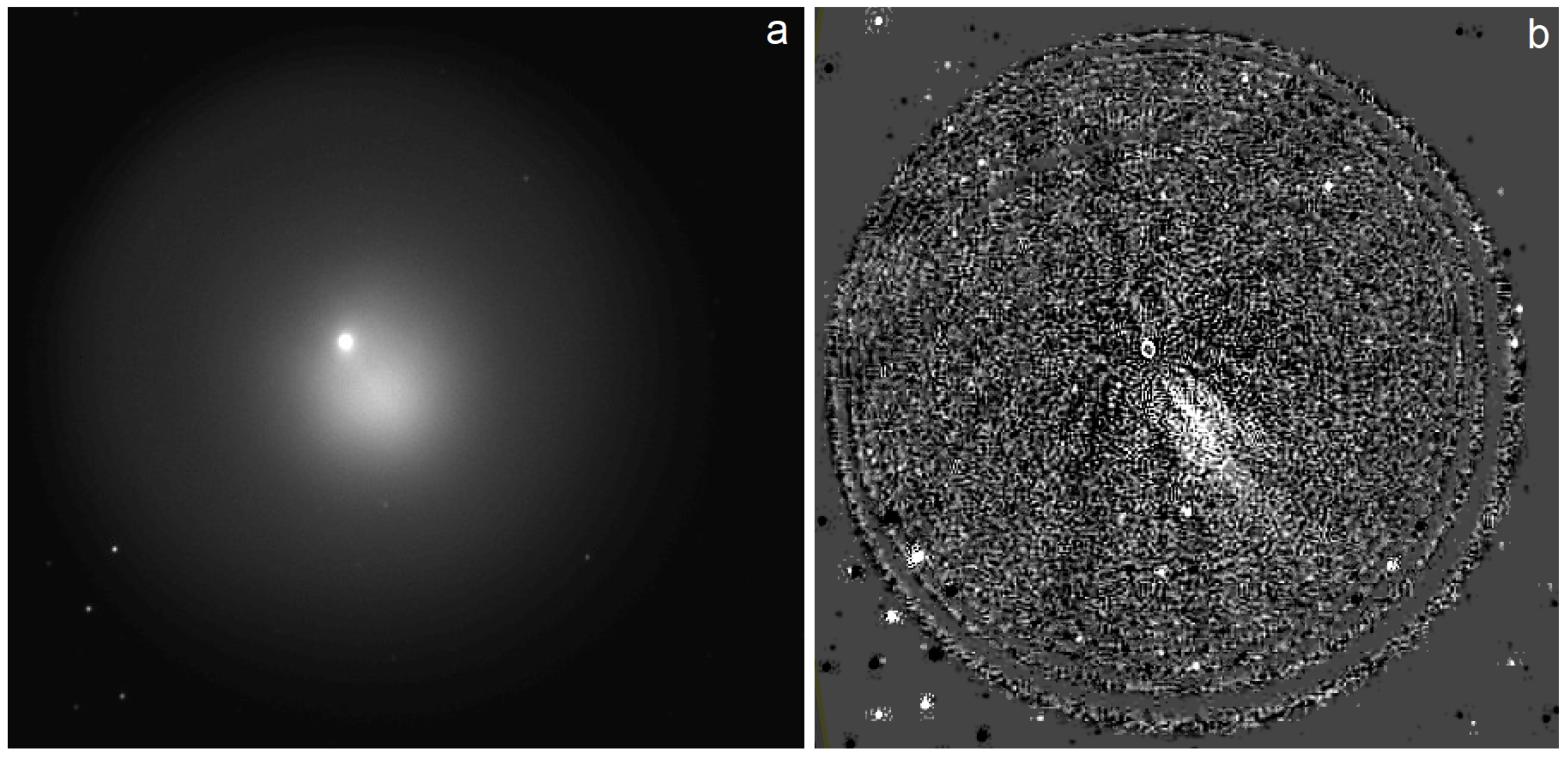}
\end{center}
\caption{Image of comet 17P/Holmes acquired on 2007 October 30.1 with the 0.8-m Jacobus Kapteyn Telescope (IAC80). (a) Overall view of the coma, showing pronounced asymmetries and an apparent diameter of approximately 7.1 arcmin ($\sim$ 500,000 km). (b) Larson–Sekanina filtered image revealing multiple expanding jets, the brightest extending from the false nucleus toward a position angle of 214$^{\circ}$, as well as a system of concentric rings. These structures are likely the result of multiple dust-producing episodes. The prominent jet and the condensation visible below it support a scenario involving the massive release of large fragments that subsequently fragmented within the coma, giving rise to the observed layered morphology.}
\label{F2}
\end{figure}

\begin{figure}
\begin{center}
\includegraphics[width=6cm]{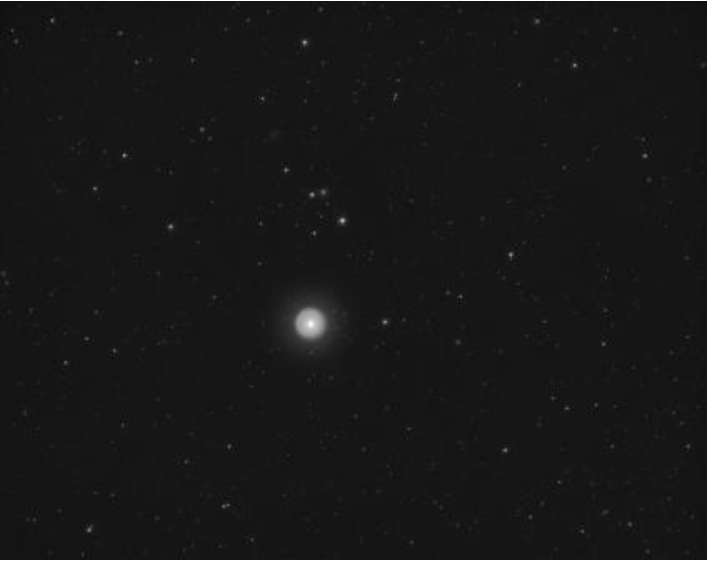}
\end{center}
\caption{Comet 17P/Holmes observed on 2007 October 28 at 22:49 UT. Image obtained using 0.07-m f/2.8 Camera lens + CCD Starlight HX916 at K02 Eastwood Observatory, Leigh-on-Sea, Essex, UK. 24 $\times$ 10-second exposures through UV/IR cut filter. Orientation = N up and E to the left. Field of view = 165' $\times$ 130'. Image scale 7.6'' per pixel.}
\label{F2a}
\end{figure}

\begin{figure}
\begin{center}
\includegraphics[width=6cm]{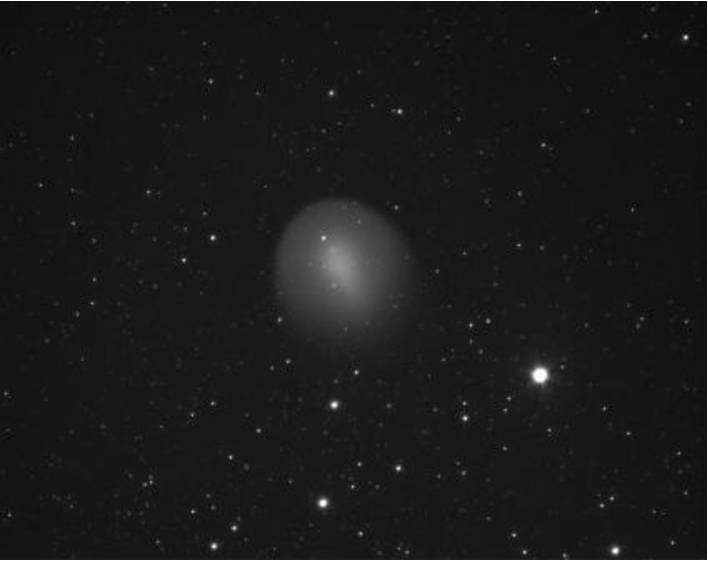}
\end{center}
\caption{Comet 17P/Holmes observed on 2007 November 15 at 20:09 UT. Image obtained using 0.07-m f/2.8 Camera lens + CCD Starlight HX916 at K02 Eastwood Observatory, Leigh-on-Sea, Essex UK. 10 $\times$ 60-second exposures through UV/IR cut filter. Orientation = N up and E to the left. Field of view = 165' $\times$ 130'. Image scale 7.6'' per pixel.}
\label{F2b}
\end{figure}

\begin{figure}
\begin{center}
\includegraphics[width=9cm]{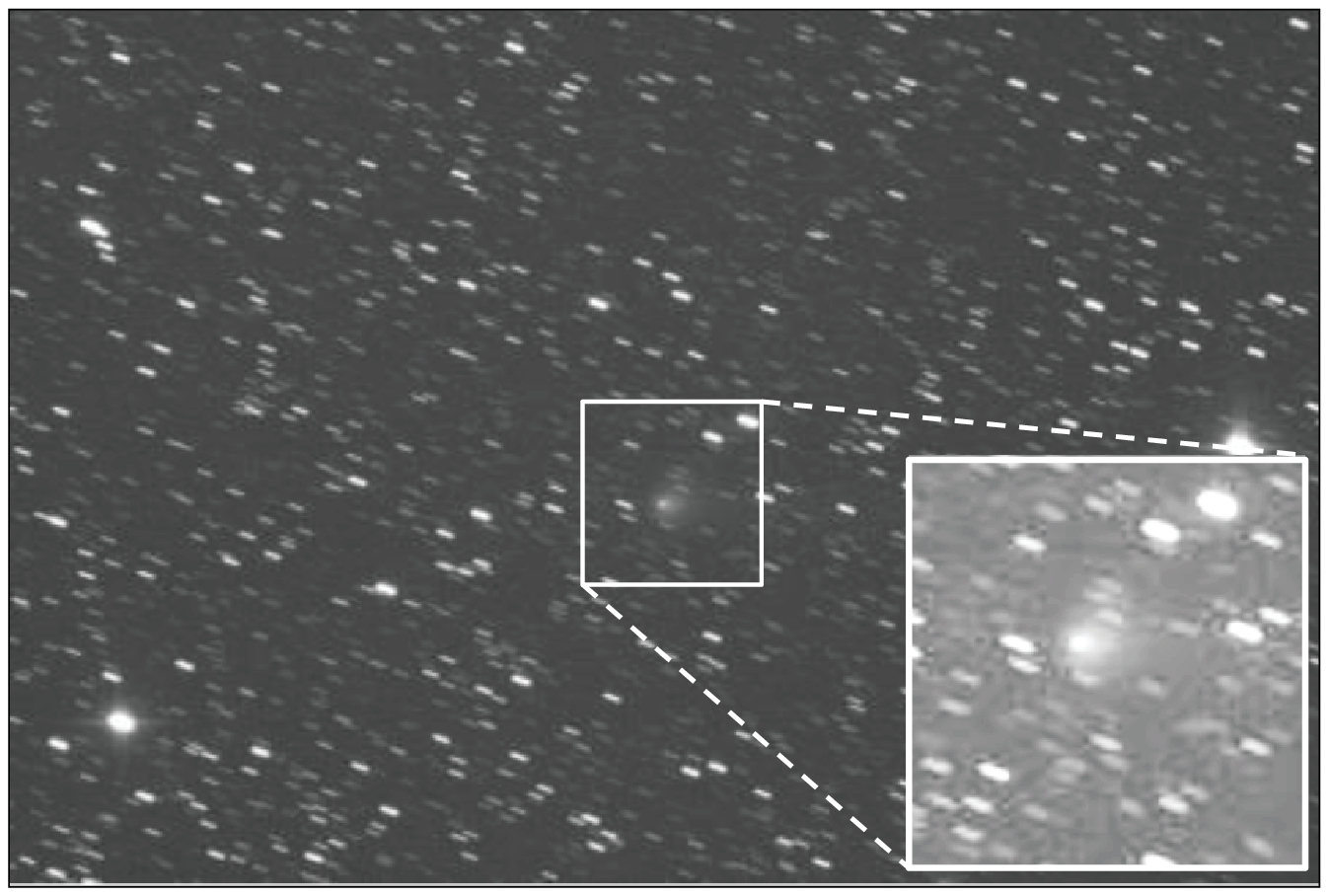}
\end{center}
\caption{Comet 17P/Holmes observed on 2021 August 10 at 03:18 UT. Image obtained remotely using 0.32-m f/8 Corrected Dall-Kirkham astrograph + CMOS QHY600M at Z10 PGC Fregenal de la Sierra observatory, Extremadura, Spain. 20 $\times$ 60-second exposures through UV/IR cut filter. m$_{1}$=13.9 against Gaia DR2 G in 45.1'' rad aperture m$_{2}$=16.4 against Gaia DR2 G in 5.1'' rad aperture. Orientation = N up and E to the left. Field of view = 30.6' $\times$ 20.4' Image scale 1.22'' per pixel.}
\label{F2c}
\end{figure}

\begin{figure}
\begin{center}
\includegraphics[width=9cm]{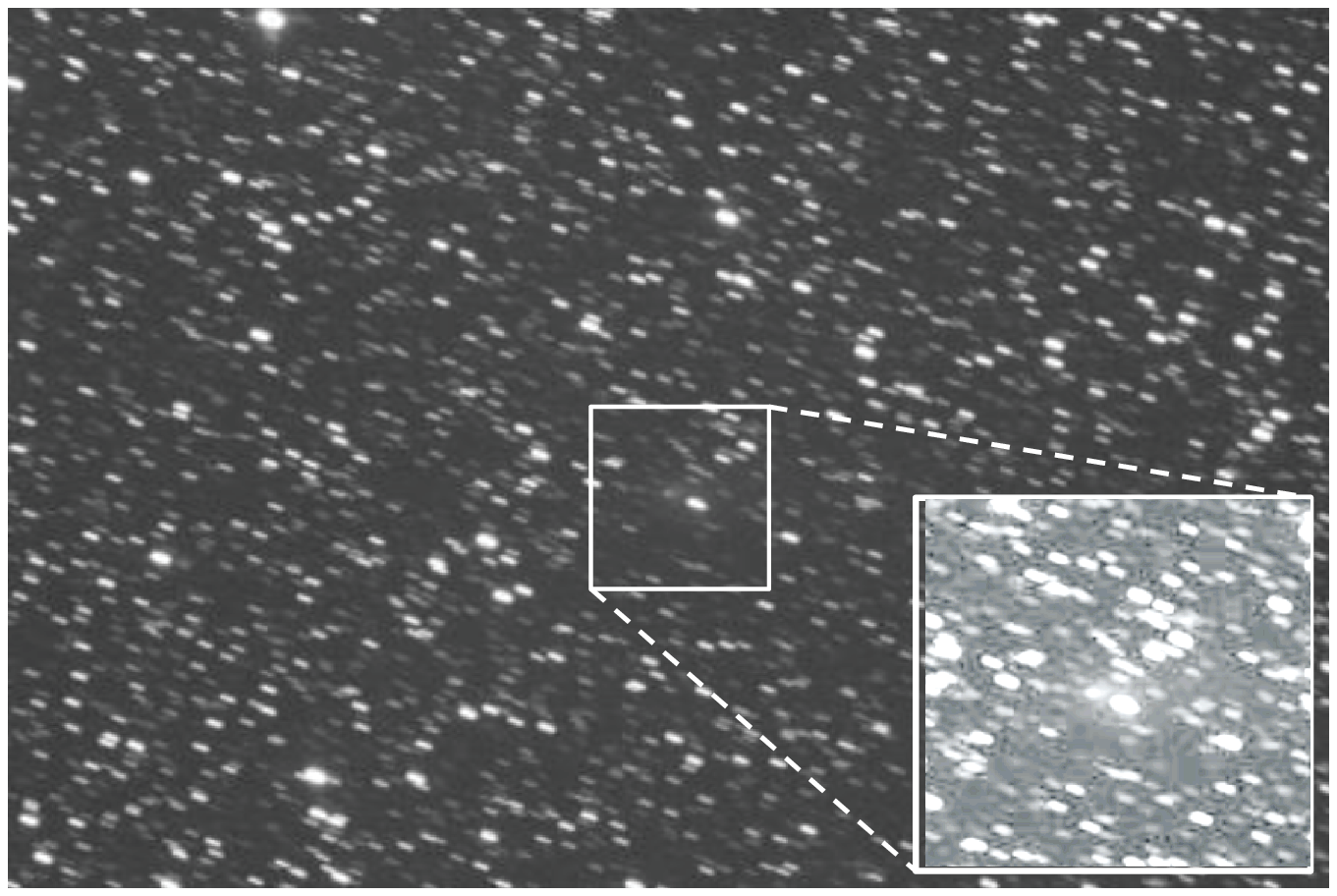}
\end{center}
\caption{Comet 17P/Holmes observed on 2021 August 17 at 02:59 UT. Image obtained remotely using 0.32-m f/8 Corrected Dall-Kirkham astrograph + CMOS QHY600M at Z10 PGC Fregenal de la Sierra observatory, Extremadura, Spain. 19 $\times$ 60 second exposures through UV/IR cut filter. m$_{2}$=17.2 against Gaia DR2 G in 5.1'' rad aperture. Orientation = N up and E to the left. Field of view = 30.6' $\times$ 20.4'. Image scale 1.22'' per pixel.}
\label{F2d}
\end{figure}

\begin{figure}
\begin{center}
\includegraphics[width=9cm]{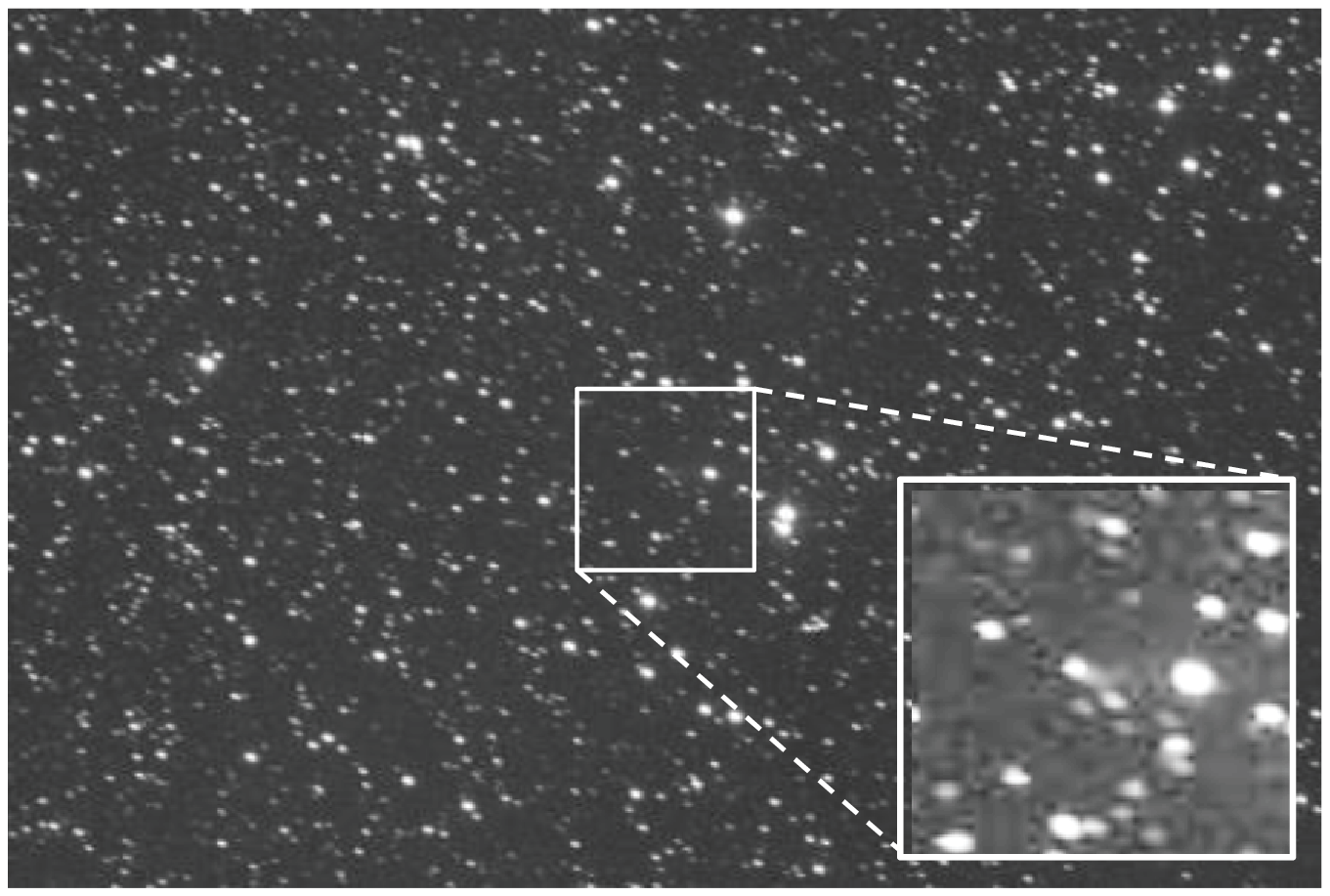}
\end{center}
\caption{Comet 17P/Holmes observed on 2021 August 21 at 03:03 UT. Image obtained remotely using 0.32-m f/8 Corrected Dall-Kirkham astrograph + CMOS QHY600M at Z10 PGC Fregenal de la Sierra observatory, Extremadura, Spain. 19 $\times$ 60 second exposures through UV/IR cut filter. m$_{2}$=17.5 against Gaia DR2 G in 5.1'' rad aperture. Orientation = N up and E to the left. Field of view = 30.6' $\times$ 20.4'. Image scale 1.22” per pixel.}
\label{F2e}
\end{figure}

Since 1964, 17P has been observed at every perihelion passage, typically appearing as an inconspicuous telescopic object until the extraordinary mega-outburst of 2007. To date, the observational record documents ten distinct outbursts since the comet’s discovery, with brightness variations ranging from 0.4 to 14 magnitudes (Tab.~\ref{Table_1a} and Figs.~\ref{F2}--\ref{F2e}). The 2007 mega-outburst event remains the most extreme cometary outburst ever recorded, exhibiting a brightness increase of $\Delta m = 14$ magnitudes. 

\begin{table*}
\caption{Amplitudes of the outbursts of comet 17P/Holmes from 1892 to 2021.}
\label{Table_1a}
\begin{tabular}{ccccc}
\hline
Date of the outburst & Heliocentric distance  & Outburst amplitude & Outburst Class as defined & Reference \\ & of the outburst [au] & [magnitude] & in \citep{Gritsevich2025b}\\
 \hline
November 4, 1892 & 2.38 & 9.0 - 10.0 & Strong outburst & \cite{Altenhoff2009,Miles2015}\\
January 16, 1893 & 2.64 & 4.0 - 6.0 & Intense outburst &  \cite{Altenhoff2009,Miles2015}\\
July 4, 1899 & 2.19 & 3.0 - 4.0 & Typical outburst &  \cite{Miles2015}\\
October 23.3 $\pm$ 0.3, 2007 & 2.47 & 14.0 & Mega-outburst & \cite{Lin2009,Montalto2008}\\ 
January 4, 2009 & 4.17 & 0.9 & Glow variation & \cite{Miles2015}\\
May 10, 2012 & 4.40 & 1.3 & Mini-outburst & \cite{Miles2015}\\
November 8, 2014 & 2.65 & 0.6 - 0.7 & Glow variation & \cite{Miles2015}\\
January 26, 2015 &  2.99 & 3.5 - 4.0 & Typical outburst & \cite{Miles2015}\\
July 1, 2021 & 2.32 & 0.4 & Glow variation & This work\\
August 3, 2021 & 2.42 & 1.4 & Mini-outburst & This work\\
\hline 
\end{tabular}
\end{table*}

The 2007 outburst occurred on October 23.3 $\pm$ 0.3 UT, when the comet was at a heliocentric distance of 2.47~au \citep{Montalto2008,Hsieh2010,Trigo2008,Reach2010,Li2011,Wesołowski2018,Gritsevich2022,Nissinen2025}. \cite{Stevenson2012} reported that, approximately three weeks after the mega-outburst, on 2007 November 12, the comet appeared brighter than expected, which would be consistent with a second outburst occurring at the nucleus shortly before their observations. However, such a variation is not clearly evident in the numerous follow-up observations of 17P, including photometric measurements from the Comet Observation Database (COBS; https://www.cobs.si/). A substantial number of observations exist between 2007 October 23 and early March 2008, when the comet entered solar conjunction. Within these data, no clear signature of a November 2007 outburst is apparent above the level of the observational scatter.

Despite extensive observational and theoretical investigations of the 2007 outburst, several key physical properties of the ejected material remain poorly constrained. In particular, observations show that simple sunward-directed or spherically symmetric velocity fields \citep{Gritsevich2022} are insufficient to reproduce the full complexity of the observed expansion, implying a broad, evolving distribution of particle ejection velocities. Because the subsequent dynamical evolution of the ejecta depends sensitively on particle size, mass, and coupling to sublimating gas, robust constraints on the particle size distribution are essential for interpreting both the brightness evolution of the coma and the long-term fate of the released material.

The size and number distributions of dust particles produced during outbursts constitute a major source of uncertainty in models of cometary activity. These parameters directly govern the total mass released, the efficiency of light scattering within the coma, and the fraction of ejecta capable of surviving radiation pressure and planetary perturbations to form coherent dust trails. In general, large outbursts may play a significant role in supplying meteoroid populations and, under favourable dynamical conditions, can contribute to the formation or enhancement of meteor streams when Earth-crossing orbits are established. Although 17P itself is not currently associated with an Earth-intersecting meteoroid stream, it provides an important benchmark for assessing the efficiency with which episodic cometary activity injects solid material into interplanetary space.

Motivated by these considerations, we conducted a systematic analysis of observations of 17P obtained during its past returns. Using the observed brightness amplitudes of documented outbursts from 1892 to 2021, with particular emphasis on the extraordinary 2007 event, we constrain the size distribution and total mass of porous dust agglomerates ejected from 17P and provide physically motivated inputs for dust-trail modelling.


\section{Mathematical model of comet activity}
\label{A1}
Integrating high-quality observational data with refined modeling approaches allows for a better understanding of the physical processes driving outbursts, the behavior of the ejected material, and the long-term evolution of 17P. We use a recently developed numerical model that enables the determination of the mass ejected during a cometary outburst \citep{Gritsevich2025b}. Based on these constraints, we estimate the approximate number of particles responsible for scattering the incident sunlight. The results presented here focus on the 2007 mega-outburst.

Estimating the mass ejected during an outburst and determining the total number of particles in the coma is a complex task influenced by numerous parameters. Among these, one of the most important factors is the flux of water-ice sublimation occurring through the porous structure of the cometary nucleus \citep{Wesołowski2025}. To determine the ejected mass, we employ Pogson's law, which can be written as \citep{Gronkowski2009a,Gronkowski2009b,Gronkowski2018,Wesołowski2018,Wesołowski2020}:

\begin{equation}
\Delta m = m_{2} - m_{1} = -2.5 {\rm log} \left(\frac{E(t_{2})}{E(t_{1})}\right),
\label{0a}
\end{equation}
\noindent where m$_{1}$ and m$_{2}$ are the apparent magnitudes of the comet in the quiet sublimation phase (t$_{1}$) and the outburst phase (t$_{2}$). The quantities E(t$_{1}$) and E(t$_{2}$) denote the observed flux from the comet at Earth during the quiet sublimation and outburst phases, respectively. Since E(t$_{1}$) and E(t$_{2}$) are proportional to the scattering cross-sections of particles reflecting sunlight, Eq.~(\ref{0a}) can be rewritten as:
\begin{equation}
\Delta m = -2.5 {\rm log} \left(\frac{p_{\mathrm{N}} C_{\mathrm{N}} + p_{\mathrm{2}} S_{2}}{p_{\mathrm{N}} C_{\mathrm{N}} + p_{\mathrm{1}} S_{1}}\right),
\label{0b}
\end{equation}
\noindent where p$_{\mathrm{N}}$ is the phase function of the cometary nucleus, calculated using the relation given by \cite{Guttler2017} (p$_{\mathrm{N}}$ = 0.019). The parameters p$_{1}$ and p$_{2}$ denote the phase functions of particles scattering the incident sunlight and were calculated using the formula of \citep{Henyey1941}. C$_{\mathrm{N}}$ represents the scattering cross-section of the cometary nucleus, while S$_{1}$ and S$_{2}$ are the total scattering cross-sections of the ice–dust particles present in the coma during the phases (t$_{1}$) and (t$_{2}$), respectively. Since S$_{1}$=C(t$_{1}$), and S$_{2}$=C(t$_{2}$) + C$_{\mathrm{ej}}$, Pogson's law can be expressed in the form:
\begin{equation}
\Delta m = -2.5 {\rm log} \left(\frac{p_{\mathrm{N}} C_{\mathrm{N}} + p_{\mathrm{2}}\left(C(t_{2}) + C_{\mathrm{ej}}\right)}{ p_{\mathrm{N}} C_{\mathrm{N}} + p_{\mathrm{1}}C(t_{1})}\right). 
\label{0c}
\end{equation}
\noindent Here, C(t$_{1}$) is the total scattering cross-section of particles present in the coma during the quiet sublimation phase, C(t$_{2}$) is the corresponding cross-section during the outburst phase, and C$_{\mathrm{ej}}$ denotes the scattering cross-section associated with material released during the fragmentation of a part of the cometary nucleus. The individual scattering cross-sections in Eq. (\ref{0c}) are defined as follows:
\begin{equation}
C_{\mathrm{N}} = A_{\mathrm{N}} S_{\mathrm{N}},
\label{0aa}
\end{equation}
\begin{equation}
C(t_{1}) = \pi N_{1}  \int_{\mathrm{r_{\mathrm{min}}}}^{\mathrm{r_{\mathrm{max}}}}  Q_{\mathrm{scat}} r^{2-q}  dr, 
\label{0bb}
\end{equation}
\begin{equation}
C(t_{2}) = \pi N_{2}  \int_{\mathrm{r_{\mathrm{min}}}}^{\mathrm{r_{\mathrm{max}}}}  Q_{\mathrm{scat}} r^{2-q}  dr, 
\label{0cc}
\end{equation}
and 
\begin{equation}
C_{\mathrm{ej}} = \pi N_{\mathrm{ej}} \int_{\mathrm{r_{\mathrm{min}}}}^{\mathrm{r_{\mathrm{max}}}}  Q_{\mathrm{scat}} r^{2-q}  dr. 
\label{0dd}
\end{equation}
\noindent Then Pogson's law, given by Eq. (\ref{0c}), takes the following form:
\begin{equation}
\Delta m = -2.5\mathrm{log}\frac{p_{\mathrm{N}} A_{\mathrm{N}} S_{\mathrm{N}} + p_{\mathrm{2}} \pi \left(N_{2} + N_{\mathrm{ej}}\right) \int_{\mathrm{r_{\mathrm{min}}}}^{\mathrm{r_{\mathrm{max}}}} Q_{\mathrm{scat}}r^{\mathrm{2-q}}dr}{p_{\mathrm{N}} A_{\mathrm{N}} S_{\mathrm{N}} + p_{\mathrm{1}} \pi N_{1} \int_{\mathrm{r_{\mathrm{min}}}}^{\mathrm{r_{\mathrm{max}}}}  Q_{\mathrm{scat}}r^{\mathrm{2-q}}dr}.
\label{MW1}
\end{equation}
\noindent We assume that the photometric measurements correspond to stages in which the coma is predominantly optically thin, or at most only marginally optically thick, such that deviations from linearity between brightness and scattering cross-section remain negligible. Equation~(\ref{MW1}) is therefore derived under the assumption of single-scattering conditions. In this regime, the observed brightness is directly proportional to the total dust scattering cross-section, and the linear scaling adopted in our formulation is strictly valid.

During exceptionally strong outbursts, however, the innermost coma may temporarily reach optical depths of order unity at the earliest stages of expansion. Under such conditions, multiple scattering and radiative transfer effects introduce a non-linear relationship between brightness and scattering cross-section. The present formulation shall thus be regarded as an optically thin approximation, applicable to the global coma once significant expansion has occurred, while early-time inner-coma effects are not explicitly treated in this model. 

In Eq. (\ref{MW1}) A$_{\mathrm{N}}$ is the albedo of the cometary nucleus (A$_{\mathrm{N}}$ = 0.04),  S$_{\mathrm{N}}$ is the total cross-sectional area of the cometary nucleus (S$_{\mathrm{N}} \approx$ 8.245$\cdot$10$^{6}$ m$^{2}$, in these calculations it was assumed that the radius of the nucleus R$_{\mathrm{N}}$ = 1620 m, \cite{Wesołowski2018}), and Q$_{\mathrm{scat}}$ is the scattering coefficient, which depends on the refractive index of a given particle, its characteristic radius, and the wavelength of the incident light. The remaining symbols have the following meanings: N$_{1}$ and N$_{2}$ denote the number of particles that were lifted into the coma during the quiet sublimation phase and during the outburst. N$_{\mathrm{ej}}$ represents the number of particles produced by the destruction of a fragment of the cometary nucleus, $r$ is the radius of particles, and $q$ is the power-law index of the particle size distribution. For comet 17P, the particle distribution in the coma follows a power law with the index $q$ ranging from 2 to 4 \citep{Stevenson2010}. The integration limits were chosen to cover a wide range of particle radii, from 10$^{-7}$ to 10$^{-2}$ m. 

\cite{Wesołowski2020} demonstrated that the contribution of the cometary nucleus cross-section to the total scattering cross-section is negligible and can therefore be safely ignored. This allows us to focus on the scattering of incident solar radiation by dust particles released into the coma. Furthermore, a cometary outburst typically occurs over a time interval that is short compared to the orbital period. Consequently, the phase angle changes only marginally between the pre-outburst and outburst states. It is therefore reasonable to assume that the phase angles just before and during the outburst are effectively identical. Under these assumptions,  Eq. (\ref{MW1}) can be simplified and takes the form:

\begin{equation}
\Delta m = -2.5\mathrm{log}\frac{\pi \left(N_{2} + N_{\mathrm{ej}}\right) \int_{\mathrm{r_{\mathrm{min}}}}^{\mathrm{r_{\mathrm{max}}}} Q_{\mathrm{scat}}r^{\mathrm{2-q}}dr} {\pi N_{1} \int_{\mathrm{r_{\mathrm{min}}}}^{\mathrm{r_{\mathrm{max}}}}  Q_{\mathrm{scat}}r^{\mathrm{2-q}}dr},
\label{MW1aa}
\end{equation}

\noindent where the individual numbers of particles that scatter incident sunlight can be expressed as \citep{Wesołowski2020}:
\begin{equation}
N_{1} = \frac{3 \kappa _{1} \eta_{1} R_{\mathrm{N}}^{2}R_{\mathrm{c_{1}}}F_{\mathrm{i}}(T)}{\rho_{\mathrm{agl}}v_{\mathrm{g,i}}\int_{\mathrm{r_{min}}}^{\mathrm{r_{max}}} r^{3-q} dr},
\label{MW2}
\end{equation}
\begin{equation}
N_{2} = \frac{3 \kappa _{2} R_{\mathrm{N}}^{2}R_{\mathrm{c_{2}}}(\eta_{1} + \Delta\eta)F_{\mathrm{i}}(T)}{\rho_{\mathrm{agl}}v_{\mathrm{g,i}}\int_{\mathrm{r_{min}}}^{\mathrm{r_{max}}} r^{3-q} dr},
\label{MW3}
\end{equation}
\noindent and
\begin{equation}
N_{\mathrm{ej}} = \frac{3 M_{\mathrm{ej}}}{4 \pi \rho_{\mathrm{gr,x}}(1 - \psi) \int_{\mathrm{r_{min}}}^{\mathrm{r_{max}}} r^{3-q} dr}.
\label{MW4}
\end{equation}
In Eqs. (\ref{MW2}-\ref{MW4}), the individual symbols denote the following quantities: R$_{\mathrm{c_{1}}}$ is the radius of the coma in the quiet sublimation phase (R$_{\mathrm{c_{1}}}$ = 1.18$\cdot$10$^{8}$ m, \citep{Wesołowski2018}), R$_{\mathrm{c_{2}}}$ is the radius of the coma in the outburst phase (R$_{\mathrm{c_{2}}}$ = 4.64$\cdot$10$^{8}$ m, \citep{Wesołowski2018}), $\rho_{\mathrm{agl}}$ is the density of the porous agglomerate ($\rho_{\mathrm{agl}}$ = 875 kg$\cdot$m$^{-3}$, \citep{Gundlach2015}), F$_{\mathrm{i}}$ is the sublimation flux, and v$_{\mathrm{g,i}}$ is the gas velocity. The remaining symbols denote the following quantities: $\eta_{1}$ is the active surface during quiet sublimation, and $\Delta\eta$ is a correction related to the ejection of a fragment of the cometary nucleus during the outburst, which is calculated based on the following relationship:
\begin{equation}
\Delta\eta = \frac{M_{\mathrm{ej}}}{4 S_{\mathrm{N}} h \rho_{\mathrm{gr,x}}(1 - \psi)},
\end{equation}
where M$_{\mathrm{ej}}$ is the mass of the ejected material, $\rho_{\mathrm{gr,x}}$ is the density of the particles that scatter the incident sunlight (in our considerations, we take into account three types of particles: water ice ($\rho_{\mathrm{gr,ice}}$ = 933 kg$\cdot$m$^{-3}$), organic matter ($\rho_{\mathrm{gr,org}}$ = 1600 kg$\cdot$m$^{-3}$), and silicate ($\rho_{\mathrm{gr,sil}}$ = 2950 kg$\cdot$m$^{-3}$), $h$ is the thickness of the surface layer that has undergone destruction ($h$ = 10 m, \citep{Wesołowski2020}) and $\psi$ is the particle porosity ($\psi$ = 0.7, \citep{Wesołowski2021}). For the quiet sublimation, it was assumed that the dust-to-gas mass ratio is $\kappa_{1}$ = 1, while for the outburst phase this ratio is assumed to be $\kappa_{2}$ = 3 \citep{Wesołowski2024}. The value of the $\kappa_{2}$ parameter results from the rapid emission of porous agglomerates originating from the destruction of a fragment of the cometary nucleus.

The sublimation flux in Eqs. (\ref{MW2}-\ref{MW3}) governs the sublimation activity of the comet during the quiet phase and is given by: 
\begin{equation}
F = m_{\mathrm{g}} \dot{Z}, 
\label{0fg}
\end{equation}
\noindent where $m_{\mathrm{g}}$ is the mass of a gas molecule and $\dot{Z}$ is the sublimation rate of ice. To estimate the value of this flux as accurately as possible, we consider three representative cases. Hereafter, $F_i(T)$ denotes the sublimation flux corresponding to case $i = 1, 2, 3$. 

In the first case ($i = 1$), the porous agglomerates that constitute the cometary nucleus are covered by a thin layer of desiccated porous dust. In this situation, the sublimation flux $F_1$ can be expressed as \citep{Kossacki2023,Wesołowski2025}: 
\begin{equation}
F_{1}(T) = C_{\mathrm{\psi}}\alpha_{\mathrm{s}}(T) \psi_{\mathrm{d}} \left(\frac{20 + 8\frac{d}{r_{\mathrm{p}}}}{20 + 19 \frac{d}{r_{\mathrm{p}}} + 3(\frac{d}{r_{\mathrm{p}}})^{2}}\right){\left( \frac{\mu}{2 \pi R_{\mathrm{g}} T} \right)}^{0.5} p_{\mathrm{sat}}. 
\label{MW_F1}
\end{equation}
In the second case (i=2), the agglomerates are exposed and sublimation takes place directly from their surface. Then the sublimation flux $F_2$ can be expressed as: 
\begin{equation}
F_{2}(T) = C_{\mathrm{\psi}} \alpha_{\mathrm{s}}(T)  {\left( \frac{\mu}{2 \pi R_{\mathrm{g}} T} \right)}^{0.5} p_{\mathrm{sat}}.  
\label{MW_F2}
\end{equation}
In the third case (i=3), the sublimation flux was corrected by the anomalous evaporation coefficient (the value of this coefficient is in the range of 0 <$\beta$< 1). Then the sublimation flux $F_3$ can be expressed as: 
\begin{equation}
F_{3}(T) = \beta \psi {\left( \frac{\mu}{2 \pi R_{\mathrm{g}}T}\right)}^{0.5} p_{\mathrm{sat}}. 
\label{MW_F3}
\end{equation}
In Eqs. (\ref{MW_F1}-\ref{MW_F3}), the individual symbols denote the following quantities: C$_{\mathrm{\psi}}$ is a parametric function, $\psi_{\mathrm{d}}$ is the porosity of the dust layer ($\psi_{\mathrm{d}}$ = 0.45, $d$ is the thickness of the layer ($d$ = 1$\cdot$10$^{-3}$ m), r$_{\mathrm{p}}$ is the radius of the pores (r$_{\mathrm{p}}$ = 0.2$\cdot$10$^{-3}$ m), $\mu$ is the molar mass of water ice ($\mu$ = 0.018 kg$\cdot$mol$^{-1}$), R$_{\mathrm{g}}$ is the universal gas constant (R$_{\mathrm{g}}$ = 8.314 J$\cdot$mol$^{-1}\cdot$K$^{-1}$), $\beta$ is the dimensionless sticking coefficient of gas molecules onto the surface ($\beta$=0.15), and $p_{\mathrm{sat}}$ is the equilibrium pressure, which was calculated using the classical formula: 
\begin{equation}
p_{\mathrm{sat}} = 3.56 \cdot 10^{12} \, \mathrm{exp} \left(-6141.667\cdot T^{-1}\right).
\label{MW_pressure}
\end{equation}
The sublimation coefficient depends on the temperature determined from the energy balance equation given by:
\begin{equation}
\frac{S_{\odot}(1 - A_{\mathrm{N}})}{r_{\mathrm{h}}^{2}} \mathrm{max}(\mathrm{cos}\,\Theta,0) -
\epsilon \,\sigma_{\mathrm{B}} \,T_{\mathrm{i}}^{4} - H\,F_{\mathrm{i}}(T) = 0, 
\label{MW_temp}
\end{equation} 
\noindent where S$_{\odot}$ is the solar constant at 1 au (S$_{\odot}$ = 1361.1$\pm$0.5 W m$^{-2}$), $\Theta$ is the solar zenithal angle,  $\epsilon$ is the emissivity ($\epsilon$ = 0.9), $\sigma_{\mathrm{B}}$ is the Stefan–Boltzmann constant, ($\sigma_{\mathrm{B}}$ = 5.67$\cdot$10$^{-8}$ W$\cdot$m$^{-2}\cdot$K$^{-4}$), T$_{\mathrm{i}}$ is the temperature, which depends on the value of the water-ice sublimation flux F$_{\mathrm{i}}$(T), and $H$ is the latent heat of sublimation of water ice ($H$ = 2.83$\cdot$10$^{6}$ J$\cdot$kg$^{-1}$).

The calculated thermodynamic parameters used in the numerical model are presented in Table \ref{Tabdm}. These three cases represent progressively more efficient sublimation regimes, corresponding to different physical states of the surface and subsurface layers.
\begin{table}
\begin{center}
\caption{Values of temperature and sublimation flux of water ice for the three cases discussed in this paper.}
\label{Tabdm}
\begin{tabular}{ccc}
\hline
Case & Temperature (T$_{\mathrm{i}}$) [K] & Sublimation flux (F$_{\mathrm{i}}$) [kg$\cdot$m$^{-2}\cdot$s$^{-1}$]\\
 \hline
i = 1 & 233.922 & 2.165$\cdot$10$^{-5}$ \\
i = 2 & 220.613 & 3.291$\cdot$10$^{-5}$ \\
i = 3 & 204.517 & 4.405$\cdot$10$^{-5}$ \\
\hline 
\end{tabular}
\end{center}
\end{table}
\subsection{Determining the mass ejected during the outburst}
\label{3.1}
\begin{figure}
\includegraphics[width=8.5cm]{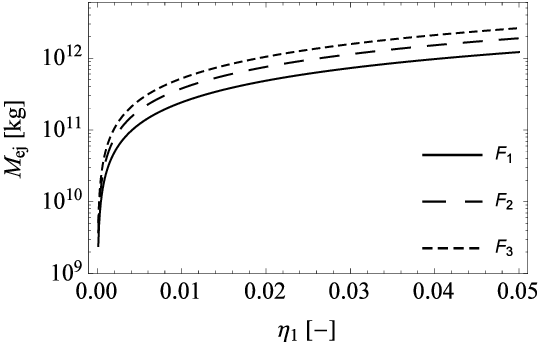}
\caption{Mass ejected during the outburst of comet 17P in 2007. The calculations assume a wide range of sublimation active surfaces (parameter $\eta$) and three values of sublimation flux. Additionally, it was assumed that the scattering of incident sunlight occurs on porous ice particles, and the cometary activity is controlled by the sublimation of water ice.}
\label{F_1}
\end{figure}

\begin{figure}
\includegraphics[width=8.5cm]{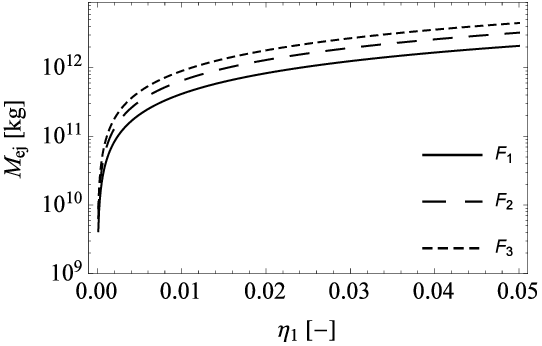}
\caption{Mass ejected during the outburst of comet 17P in 2007. The calculations assume that the scattering of incident sunlight occurs on porous organic particles. The remaining assumptions are same as in Fig. (\ref{F_1}).}
\label{F_2}
\end{figure}

\begin{figure}
\includegraphics[width=8.5cm]{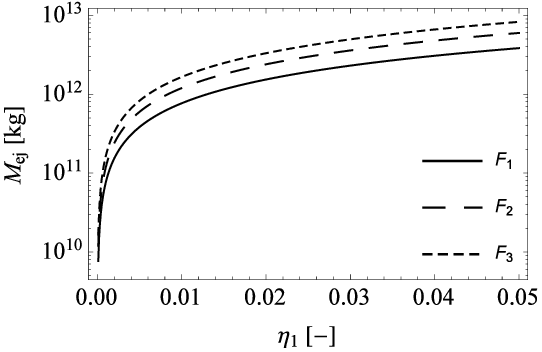}
\caption{Mass ejected during the outburst of comet 17P in 2007. The calculations assume that the scattering of incident sunlight occurs on porous dust particles. The remaining assumptions are same as in Fig. (\ref{F_1}).}
\label{F_3}
\end{figure}

\begin{figure}
\includegraphics[width=8.5cm]{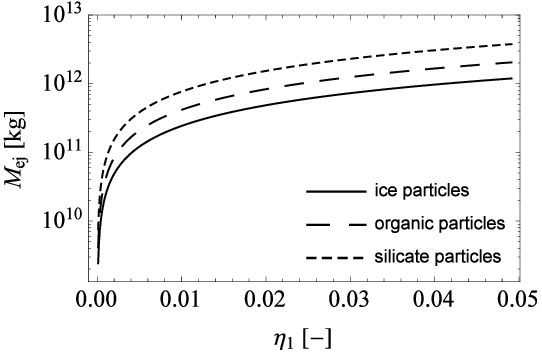}
\caption{Mass ejected during the outburst of comet 17P in 2007 as a function of the active area during quiet sublimation. Calculations include the effect of particle density on the total mass ejected, which is determined by the sublimation flux $F_1$ = 2.165$\cdot$10$^{-5}$ kg$\cdot$m$^{-2}\cdot$s$^{-1}$.}
\label{F_4}
\end{figure}

\begin{figure}
\includegraphics[width=8.5cm]{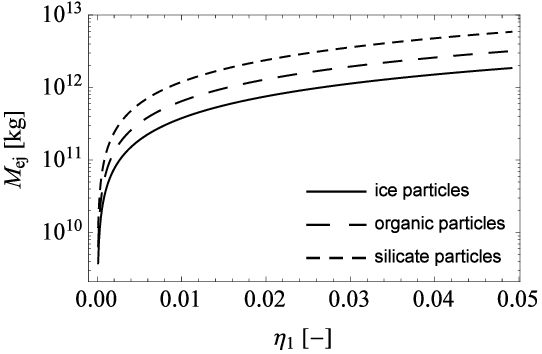}
\caption{Mass ejected during the outburst of comet 17P in 2007 as a function of the active area during quiet sublimation. Calculations include the effect of particle density on the total mass ejected, which is determined by the sublimation flux $F_2$ = 3.291$\cdot$10$^{-5}$ kg$\cdot$m$^{-2}\cdot$s$^{-1}$.}
\label{F_5}
\end{figure}

\begin{figure}
\includegraphics[width=8.5cm]{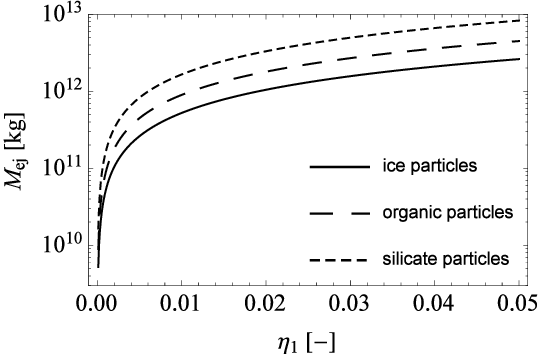}
\caption{Mass ejected during the outburst of comet 17P in 2007 as a function of the active area during quiet sublimation. Calculations include the effect of particle density on the total mass ejected, which is determined by the sublimation flux $F_3$ = 4.405$\cdot$10$^{-5}$ kg$\cdot$m$^{-2}\cdot$s$^{-1}$.}
\label{F_6} 
\end{figure}

As a result of the outburst, large amounts of gas and dust were ejected and incident sunlight scattered off the particles. Scattering by dust particles (porous agglomerates) is more effective than scattering by gas molecules. Therefore, in our numerical calculations we focus on determining the mass ejected associated with dust particles. To calculate the mass ejected from the destruction of a fragment of the cometary nucleus, Eq. (\ref{MW1}) must be solved numerically. In this solution, we use Eqs. (\ref{MW2}-\ref{MW_temp}), to analyze the influence of the water ice sublimation flux on the total mass ejected. The results of these calculations are presented in Figs. (\ref{F_1}-\ref{F_3}). Additionally, using the same set of equations, Figs. (\ref{F_4}-\ref{F_6}) show the influence of particle density on the total mass ejected during the considered outburst of comet 17P.


\noindent Based on the obtained results, the following conclusions can be reached:
\begin{itemize}
\item [1)] The ejected mass scales primarily with the fraction of the active surface during quiet sublimation (parameter $\eta$). As this active surface increases, so does the ejected mass, which ultimately determines the amplitude of the cometary outburst.
\item [2)] The smaller the value of the sublimation flux, the smaller the mass ejected. This is a consequence of the fact that a smaller flux value generates a larger outburst amplitude for the same ejected mass.
\item [3)] For porous ice agglomerates and an active surface of $\eta$ = 5\%, the ejected mass is M$_{\mathrm{ej}}$ = 1.214$\cdot$10$^{12}$ kg for flux $F_1$, 1.892$\cdot$10$^{12}$ kg for $F_2$, and 2.907$\cdot$10$^{12}$ kg for $F_3$.
\item [4)] For porous organic agglomerates and $\eta$ = 5\%, the corresponding ejected masses are 2.082$\cdot$10$^{12}$ kg ($F_1$), 3.244$\cdot$10$^{12}$ kg ($F_2$), and 4.985$\cdot$10$^{12}$ kg ($F_3$). 
\item [5)] For porous dust agglomerates and $\eta$ = 5\%, the ejected masses are 3.839$\cdot$10$^{12}$ kg ($F_1$), 5.981$\cdot$10$^{12}$ kg ($F_2$), and 9.191$\cdot$10$^{12}$ kg ($F_3$).
\item [6)] These results show that the total ejected mass increases with increasing density of porous agglomerates. This occurs because, for a given scattering cross-section constrained by observations, higher-density particles correspond to a larger total ejected mass. The mass ejected from porous dust agglomerates is 3.16 times larger than that from porous ice agglomerates and 1.84 times larger than that from porous organic agglomerates.

\item [7)] For a wide range of the parameter $\eta$ and three values of the sublimation flux, the obtained results are consistent with observational data reported by \citep{Montalto2008,Sekanina2008,Hsieh2010,Li2011,Boissier2012}.
\end{itemize}
\subsection{The number of particles in the coma}
\label{3.2}
\begin{figure}
\begin{center}
\includegraphics[width=8.5cm]{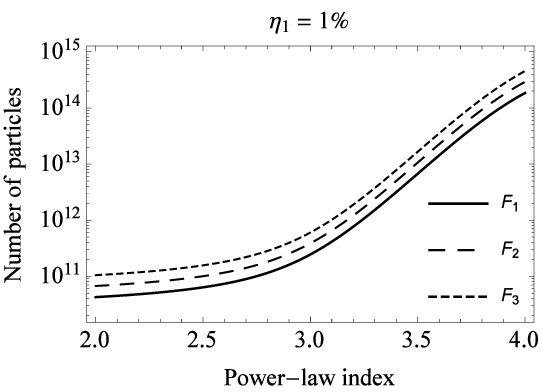}
\includegraphics[width=8.5cm]{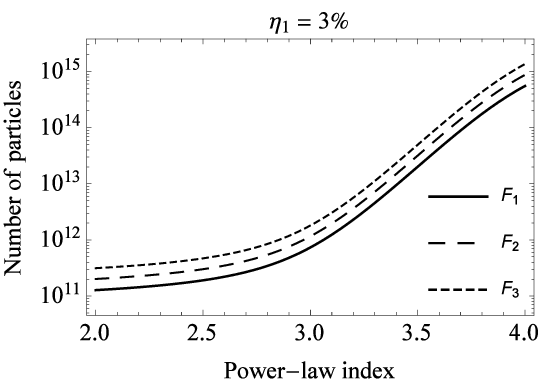}
\includegraphics[width=8.5cm]{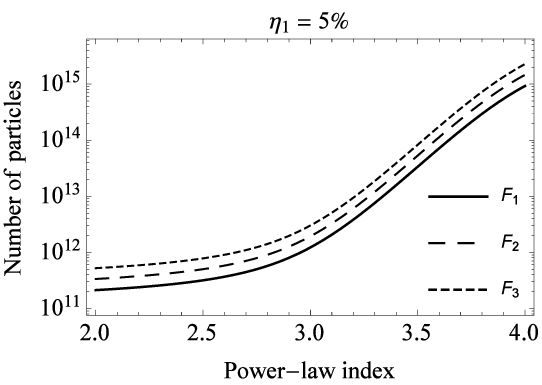}
\end{center}
\caption{The total number of particles present in the coma that scatter incident sunlight. These calculations take into account the effect of the sublimation flux and a wide range of the power-law index $q$.}
\label{F_7}
\end{figure}

Using the obtained values of the ejected mass during the outburst of comet 17P, we determine the total number of porous agglomerates scattering incident sunlight. In these calculations, the effective particle radius is derived assuming a power-law size distribution with index $q$ within the range reported by \citep{Stevenson2010}. For the adopted range of the power-law index $q$, the effective particle radius varies from 1.15 $\cdot$ 10$^{-6}$ m (for $q = 4$) to 5 $\cdot$ 10$^{-3}$ m (for $q = 2$). The resulting total number of porous agglomerates scattering incident sunlight is shown in Fig.~(\ref{F_7}).

When estimating the total number of particles present in the coma during a cometary outburst, several factors shall be taken into account:

\begin{itemize}

\item [1)] The inverse problem of determining the number of particles in the coma depends on several parameters, primarily the outburst intensity, as well as the particle size distribution, active surface fraction, and sublimation flux.
\item [2)] As the power-law index $q$ increases, the number of light-scattering particles also increases. For $q = 4$, M$_{\mathrm{ej}}$ = 2.428$\cdot$10$^{11}$ kg, $F_1$(T) = 2.165$\cdot$10$^{-5}$ kg$\cdot$m$^{-2}\cdot$s$^{-1}$, and $\eta_1 = 1\%$, the number of particles is 1.844$\cdot$10$^{14}$. For $\eta_1 = 5\%$ and M$_{\mathrm{ej}}$ = 1.214$\cdot$10$^{12}$ kg, the number increases to 9.221$\cdot$10$^{14}$. For $q = 2$, the corresponding particle numbers are 4.247$\cdot$10$^{10}$ ($\eta_1 = 1\%$) and 2.123$\cdot$10$^{11}$ ($\eta_1 = 5\%$).
\item [3)] Increasing the active surface area during the quiet sublimation phase leads to an increase in the number of particles in the coma due to the larger ejected mass and sublimation flux.
\item [4)] Increasing the sublimation flux also increases the number of particles in the coma. For $F_1$(T) = 2.165$\cdot$10$^{-5}$ kg$\cdot$m$^{-2}\cdot$s$^{-1}$ the number of particles is 1.844$\cdot$10$^{14}$, while for $F_3$(T) = 4.405$\cdot$10$^{-5}$ kg$\cdot$m$^{-2}\cdot$s$^{-1}$ it is 4.569$\cdot$10$^{14}$ (for $q = 4$ and $\eta_1 = 1\%$).
\item [5)] Since the observed brightness constrains the total scattering cross-section, variations in particle density affect the inferred mass but not the total number of particles. This follows from a scaling property of the model: both the total ejected mass and the particle mass scale linearly with density, causing the density term to cancel when computing the particle number.

\end{itemize}

\section{Discussion}
We have constrained the size distribution and total mass of porous agglomerates ejected during the 2007 outburst of 17P, providing physically motivated parameters suitable for both observational interpretation and dust-trail modelling. The ejecta consist of ice, organics, and dust, with characteristic particle sizes spanning $\sim 10^{-6}$ to $\sim 5 \times 10^{-3}\,\mathrm{m}$, depending on the assumed power-law index $q$. Steeper size distributions ($q = 4$) yield smaller mean particle sizes, whereas shallower distributions ($q = 2$) favor larger particles. Accounting self-consistently for particle size, sublimation flux, and bulk density, we find that the total number of ejected particles increases with both $q$ and sublimation flux.

These results place quantitative constraints on the physical nature of material released during large cometary outbursts and establish a direct connection between observed brightness amplitudes and the underlying particle population. In particular, the strong sensitivity of the particle number to the power-law index confirms that assumptions about the size distribution remain one of the dominant sources of uncertainty in modelling cometary comae and dust trails. Even modest variations in $q$ can produce order-of-magnitude differences in particle number for a fixed total ejected mass, with corresponding implications for the effective scattering cross-section and the observed photometric evolution.

An important implication concerns the role of sublimation-driven processes in regulating the efficiency of particle release. For a given outburst amplitude, lower sublimation fluxes require a larger total ejected mass to reproduce the observed brightness increase, whereas higher fluxes allow comparable photometric signatures to be achieved with fewer particles. It is therefore necessary to incorporate physically motivated sublimation models when interpreting outburst observations, rather than relying on simple empirical brightness-mass conversions. The range of sublimation fluxes explored here spans physically plausible surface conditions on 17P and demonstrates that the 2007 mega-outburst can be reproduced without invoking extreme or non-physical parameters.

The particle size constraints derived in this study lie in the submicron-to-millimetre range and are consistent with the particle sizes employed in current long-term dust-trail modelling of 17P (see Table 1 of 
\citep{Gritsevich2022}). The particles are sufficiently small that their dynamics is strongly influenced by radiation pressure and the solar wind; such behavior is already treated self-consistently in existing dynamical models 
\citep{Grun2007,Kreslavsky2021}. The inferred sizes cannot be significantly increased without violating the physical constraints imposed by sublimation-driven release, making them a robust and physically meaningful outcome of the outburst model.

The present results primarily constrain the fine component produced directly by sublimation during outbursts. The abundance, size range, and initial velocities of larger particles that may be mechanically released during outbursts remain uncertain. Such material, if present, could also contribute to long-lived dust trails and may require separate modelling using different release mechanisms and ejection velocities. Consequently, future dust-trail simulations may benefit from exploring additional, larger populations alongside the size range constrained in our work.

Despite these uncertainties, the size constraints presented in this work provide essential, empirically grounded initial conditions for advanced dust-trail modeling. They eliminate the need for ad hoc assumptions regarding the properties of outburst-generated dust and ensure physical consistency between photometric outburst modeling and long-term dynamical simulations. Updated dust-trail models incorporating these constraints will enable more reliable assessments of the contribution of episodic outbursts to cometary dust trails and meteoroid reservoirs.

To facilitate the application of our results in dust-trail simulations, we classify the ejecta into three particle populations, following and extending the framework used in \citep{Gritsevich2022} for modelling the long-term evolution of the 17P dust trail:
\begin{itemize}
\item a fine-grained population, dominated by submicron particles produced preferentially for steep size distributions ($q \gtrsim 3.5$), which contribute strongly to the initial brightness increase but are rapidly dispersed by radiation pressure;
\item an intermediate population, comprising particles near the mean size of the distribution, which governs the short- to medium-term evolution of the coma and inner dust cloud;
\item a coarse population, consisting of the largest particles permitted by the size distribution, which carry a substantial fraction of the total mass and are most likely to survive radiation pressure and planetary perturbations to form coherent dust trails.
\end{itemize}

These population-based constraints bridge photometric outburst modelling and dynamical dust-trail simulations. The derived size ranges and particle numbers supply physically grounded initial conditions for numerical models that track the spatial and temporal evolution of ejecta over timescales of years to decades. The consistency between our results and the dust-trail modelling framework \citep{Gritsevich2022} supports the conclusion that a substantial fraction of material released during outbursts persists in interplanetary space and can be observed in the form of dust trails. The predicted particle populations offer a direct link between computed dust-trail properties and their expected surface brightness, enabling comparisons with observations.

More broadly, episodic outbursts represent an efficient mechanism for injecting solid material into the Solar System. While steady-state activity dominates the long-term mass loss of many comets, rare but extreme events such as the 2007 outburst of 17P can dominate dust production on short timescales and may play a disproportionate role in shaping dust trails and meteoroid reservoirs. Quantifying both the size distribution and total mass of outburst ejecta is therefore essential for evaluating the cumulative contribution of such events to the interplanetary dust complex.

An additional source of uncertainty arises from the intrinsic diversity of dust aggregates emitted by different comets \citep{Trigo2021}. Meteoroid streams associated with different comets exhibit markedly different mass indices and flux characteristics, implying substantial variations in aggregate structure, size distribution, and mechanical strength. Nevertheless, the approach presented in this study is readily transferable to other outbursting comets. By combining observed brightness amplitudes with physically motivated models of sublimation and particle properties, comparable constraints on ejecta populations can be derived across different cometary classes, enabling broader assessments of the role of episodic activity in cometary evolution and meteoroid-stream formation.

\section{Conclusions}

The physical properties of material released during outbursts of 17P have been constrained by linking observed brightness amplitudes to a physically motivated model of sublimation-driven dust ejection. The derived particle size distributions, characterized by power-law indices in the range $q$ = 2--4, imply characteristic particle sizes from submicron to millimetre scales and confirm that assumptions about the size distribution play a critical role in the interpretation of cometary outbursts.

Our results show that the particle population generated during an outburst is governed primarily by the interplay between the size distribution and the sublimation regime. While steeper size distributions and higher sublimation fluxes lead to larger numbers of released particles, the total ejected mass is controlled mainly by the active surface fraction and the bulk properties of the material. For the 2007 mega-outburst, the inferred ejecta masses fall in the range $10^{10}$--$10^{12}$~kg, consistent with estimates by \citep{Montalto2008,Hsieh2010}. 

The inferred ejected mass ranges were used to estimate the number of particles contributing to the scattering of incident sunlight. The particle number is one of the defining parameters for interpreting observed brightness amplitudes: coma brightness is governed not simply by the total ejected mass, but primarily by the effective scattering cross-section, which depends sensitively on both the number of particles and their size distribution. Increasing the power-law index $q$ strongly enhances the abundance of small particles, leading to a rapid growth of the total scattering area and, consequently, to a pronounced increase in brightness even for comparable ejected masses.

Particle numbers reaching $10^{11}$--$10^{15}$ indicate that the photometric signature of an outburst is dominated by a fine-grained particle population, particularly during the early stages of coma expansion. This demonstrates that modelling cometary outburst light curves requires explicit consideration of particle number and size distribution, rather than reliance on mass-based estimates alone. The weak dependence of particle number on bulk density further implies that geometric factors and size distribution, rather than material composition, primarily control the efficiency of light scattering in the coma.

As a result, the number of particles scattering sunlight provides a direct link between the observed outburst amplitude and the physical conditions on the cometary nucleus, including the sublimation rate and the fraction of the active surface. This supports the interpretation that extreme brightening, such as the 2007 outburst of 17P, arise naturally from the production of vast numbers of small particles, rather than necessarily from an exceptionally large total ejected mass.

From a dynamical perspective, incorporating the population-based ejecta constraints derived in this work into long-term dust-trail simulations will enable more realistic predictions of trail morphology, density, and observability. This is particularly important for evaluating the survivability of particles over multiple orbital revolutions of the dust trail and for identifying conditions under which outburst-generated material may evolve into detectable meteoroid streams. Future space-based observations and in situ dust measurements, including next-generation dust analysers, offer the prospect of directly sampling particles whose origin can be traced back to specific cometary outbursts.

A key next step is the direct comparison between the predicted properties of dust trails derived from our model and their observable surface brightness. By coupling computed particle size distributions and number densities with radiative scattering models, dynamical dust-trail simulations can be translated into quantitative brightness predictions. Such comparisons provide an observational test of the adopted ejecta parameters and will place strong constraints on particle albedo, bulk density, and size distribution, thereby refining dust-trail models for both individual outbursts and long-term cometary activity.

\section*{Acknowledgements}
This work was supported by the Spanish Ministry of Science, Innovation and Universities projects Nos PID2023-151905OB-I00, PID2020-118491GB-I00, Academy of Finland project no. 325806 (PlanetS), Centre for Innovation and Transfer of Natural Sciences and Engineering Knowledge, University of Rzeszów, Poland (RPPK.01.03.00-18-001/10-00), Junta de Andalucía grant P20\_010168, and the Centro de Excelencia Severo Ochoa grant CEX2021-001131-S funded by MCIN/AEI/10.13039/501100011033. The programme of development within Priority-2030 is acknowledged for supporting the research at UrFU. JMT-R work was funded by the Spanish PID2021-128062NB-I00 research grant.\\

\section*{Data Availability}

The data underlying this study are included in this article. Additional long-term observations of comet 17P/Holmes and its dust trail, formed during the 2007 outburst, are available from the authors upon reasonable request.



\begin{thebibliography}{}
\bibitem[Altenhoff et al.(2009)]{Altenhoff2009}
Altenhoff, W. J., Kreysa, E., Menten, K. M., Sievers, A., Thum, C., Weiss, A., 2009. Why did Comet 17P/Holmes burst out? A\&A 495, 975 - 978. DOI: 10.1051/0004-6361:200810458
\bibitem[Boissier et al.(2012)]{Boissier2012}
Boissier, J., Bockel\'ee-Morvan, D., Biver, N., Crovisier, J., Lellouch, E., et al., 2012. Interferometric mapping of the 3.3-mm continuum emission of comet 17P/Holmes after its 2007 outburst. Astronomy \& Astrophysics 542, A73. DOI: 10.1051/0004-6361/201219001
\bibitem[Belousov and Pavlov(2024a)]{Belousov2024a}
Belousov, D. V., Pavlov, A. K., 2024a. Cometary outbursts in the Oort cloud. Icarus 415, 116066. DOI: 10.1016/j.icarus.2024.116066
\bibitem[Belousov and Pavlov(2024b)]{Belousov2024b}
Belousov, D. V., Pavlov, A. K., 2024b. Non-gravitational Mechanism of Comets' Ejection from the Oort Cloud Due to Cometary Outbursts. Solar System Research 57, 629-635. DOI: 10.1134/S0038094623060023 
\bibitem[Filonenko and Churyumov(2006)]{Filonenko2006}
Filonenko, V. S., Churyumov, K. I., 2006. New peculiarities of cometary outburst activity. Advances in Space Research 38, 1940-1945. DOI: 10.1016/j.asr.2006.04.028
\bibitem[Gronkowski et al.(2009a)]{Gronkowski2009a}
Gronkowski, P., 2009a.The destruction of cometary grains and changes in the luminosity of comets. Astronomische Nachrichten, 330, 784. DOI: 10.1002/asna.200711234 
\bibitem[Gronkowski et al.(2009b)]{Gronkowski2009b}
Gronkowski, P., 2009b. Large cometary grains - their destruction and changes in the luminosity of comets. Monthly Notices of the Royal Astronomical Society 397, 883-889. DOI: 10.1111/j.1365-2966.2009.14994.x 
\bibitem[Gronkowski and Sacharczuk(2010)]{Gronkowski2010}
Gronkowski, P., Sacharczuk, Z., 2010. Cometary outbursts - a search for a cause of the comet 17P/Holmes outburst. Monthly Notices of the Royal Astronomical Society 408, 1207-1215. DOI: 10.1111/j.1365-2966.2010.17194.x 
\bibitem[Gronkowski et al.(2018)]{Gronkowski2018}
Gronkowski, P., Tralle, I., Weso\l{}owski, M., 2018. Visibility of comets during their outbursts and the night sky light pollution-Use the Bortle scale. Astronomische Nachrichten 339, 37-45. DOI: 10.1002/asna.201713387
\bibitem[Gritsevich et al.(2022)]{Gritsevich2022}
Gritsevich, M.,  Nissinen, M., Oksanen, A., Suomela, J., Silber, E. A., 2022. Evolution of the dust trail of comet 17P/Holmes. Monthly Notices of the Royal Astronomical Society  513, 2201-2214. DOI: 10.1093/mnras/stac822
\bibitem[Gritsevich et al.(2025a)]{Gritsevich2025a}
Gritsevich, M., Nissinen, M., Ryske, J., et al. 2025a. \textit{Rev. Mex. Astron. Astrof. Conf. Ser}. Cometary outbursts and evolution of ejected particles. In: M. Gritsevich, A. J. Castro-Tirado, P. Kubanek, S. B. Pandey,  D. Hiriart (Eds.), Revista Mexicana de Astronomia y Astrofisica: Serie de Conferencias, 59, pp. 93–100.
\bibitem[Gritsevich et al.(2025b)]{Gritsevich2025b}
Gritsevich, M., Weso\l{}owski, M., Castro-Tirado, A. J., 2025b, Mass of particles released by comet 12P/Pons-Brooks during 2023-2024 outbursts. Monthly Notices of the Royal Astronomical Society 538, 470–479. DOI: 10.1093/mnras/staf219
\bibitem[Gr\"un (2007)]{Grun2007}
Gr\"un, E., 2007. CHAPTER 34 - Solar System Dust. Encyclopedia of the Solar System (Second Edition), 621-636. 
\bibitem[Gundlach et al.(2015)]{Gundlach2015}
Gundlach, B., Blum, J., Keller, H. U., Skorov, Y. V., 2015. What drives the dust activity of comet 67P/Churyumov-Gerasimenko? Astronomy \& Astrophysics 583, A12. DOI: 10.1051/0004-6361/201525828
\bibitem[G\"uttler et al.(2017)]{Guttler2017}
G\"uttler, C., Hasselmann, P. H., Li, Y., Fulle, M., Tubiana, C., et al. 2017. Characterization of dust aggregates in the vicinity of the Rosetta spacecraft. Monthly Notices of the Royal Astronomical Society 469, S312-S320. DOI: 10.1093/mnras/stx1692
\bibitem[Hajra et al.(2017)]{Hajra2017}
Hajra, R., Henri, P., Valli$\grave{\mathrm{e}}$res, X., Galand, M., H$\acute{\mathrm{e}}$ritier, K., et al. 2017. Impact of a cometary outburst on its ionosphere. Rosetta Plasma Consortium observations of the outburst exhibited by comet 67P/Churyumov-Gerasimenko on 19 February 2016. Astronomy \& Astrophysics 607, A34. DOI:  10.1051/0004-6361/201730591
\bibitem[Henyey and Greenstein(1941)]{Henyey1941}
Henyey, L. G., Greenstein, J. L., 1941. Diffuse radiation in the Galaxy. Astrophysical Journal 93, 70-83. DOI: 10.1086/144246 
\bibitem[Holmes(1892)]{Holmes1892}
Holmes, E., 1892. Discovery of a new comet in Andromeda. The Observatory 15, 441-443. 
\bibitem[Hsieh et al.(2010)]{Hsieh2010}
Hsieh, H. H., Fitzsimmons, A., Joshi, Y., Christian, D., Pollacco, D.L., 2010. SuperWASP observations of the 2007 outburst of Comet 17P/Holmes. Monthly Notices of the Royal Astronomical Society 407, 1784-1800. DOI:  
10.1111/j.1365-2966.2010.17016.x
\bibitem[Hughes(1990)]{Hughes1990}
Hughes, D. W., 1990. Cometary outbursts - A review. Royal Astronomical Society, Quarterly Journal 31, 69-94.
\bibitem[Kelley et al.(2021)]{Kelley2021}
Kelley, M. S. P., Farnham, T. L. ,  Li, J.-Y., Bodewits, D., Snodgrass, C., et al. 2021. Six Outbursts of Comet 46P/Wirtanen. The Planetary Science Journal 2, 131. DOI:  10.3847/PSJ/abfe11
\bibitem[Kelley et al.(2022)]{Kelley2022}
Kelley, M. S. P., Kokotanekova, R., Holt, C. E., Protopapa, S., Bodewits, D., et al. 20222. A Look at Outbursts of Comet C/2014 UN$_{271}$ (Bernardinelli-Bernstein) near 20 au. The Astrophysical Journal Letters 933, L44. DOI:  10.3847/2041-8213/ac7bec 
\bibitem[Kossacki et al.(2023)]{Kossacki2023}
Kossacki, K. J., Weso\l{}owski, M., Szutowicz, S., Miko\l{}ajk\'ow, T., 2023. Outgassing of ice agglomerates. Icarus, 398, 115518. DOI:  
10.1016/j.icarus.2023.115518
\bibitem[Kr\"uger et al.(2024)]{Kruger2024}
Kr\"uger, H., Strub, P., Sommer, M., Moragas-Klostermeyer, G., Sterken, V. J., et al., 2024. Modeling the interstellar dust detections by DESTINY+ I: Instrumental constraints and detectability of organic compounds. Planetary and Space Science 254, 106010. DOI:  10.1016/j.pss.2024.106010
\bibitem[Kreslavsky et al.(2021)]{Kreslavsky2021}
Kreslavsky, M. A., Zharkova, A. Yu., Head, J. W., Gritsevich, M. I., 2021. Boulders on Mercury.  Icarus 369, 114628. DOI: 10.1016/j.icarus.2021.114628
\bibitem[Lacerda and Jewitt(2012)]{Lacerda2012}
Lacerda, P. Jewitt, D., 2012. Extinction in the Coma of Comet 17P/Holmes. Astrophysical Journal Letters 760, L2. DOI: 10.1088/2041-8205/760/1/L2
\bibitem[Li et al.(2011)]{Li2011}
Li, J., Jewitt, D., Clover, J. M., Jackson, B. V., 2011. Outburst of Comet 17P/Holmes Observed with the Solar Mass Ejection Imager. Astrophysical Journal 728, 31. DOI:  
10.1088/0004-637X/728/1/31
\bibitem[Lin et al.(2009)]{Lin2009}
Lin, Z.-Y., Lin, C.-S., Ip, W.-H., Lara, L. M., 2009. The Outburst of Comet 17p/Holmes. The Astronomical Journal 138, 625-632. DOI:  
10.1088/0004-6256/138/2/625
\bibitem[Miles(2015)]{Miles2015}
Miles, R., 2015. Four cometary outbursts in 30 days. Journal of the British Astronomical Association 125, 116–117.
\bibitem[Miles(2016a)]{Miles2016a}
Miles, R., 2016a. Discrete sources of cryovolcanism on the nucleus of Comet 29P/Schwassmann-Wachmann and their origin. Icarus 272, 387-413. DOI: 10.1016/j.icarus.2015.11.011
\bibitem[Miles(2016b)]{Miles2016b}
Miles, R., 2016b. Heat of solution: A new source of thermal energy in the subsurface of cometary nuclei and the gas-exsolution mechanism driving outbursts of Comet 29P/Schwassmann-Wachmann and other comets. Icarus 272, 356-386. DOI: 10.1016/j.icarus.2015.12.053
\bibitem[Montalto et al.(2008)]{Montalto2008}
Montalto, M., Riffeser, A., Hopp, U., Wilke, S., Carraro, G., 2008. The comet 17P/Holmes 2007 outburst: the early motion of the outburst material. Astronomy \& Astrophysics 479, L45. DOI: 10.1051/0004-6361:20079130
\bibitem[Moreno et al.(2008)]{Moreno2008}
Moreno, F., Ortiz, J. L., Santos-Sanz, P., et al., 2008. A Model of the Early Evolution of the 2007 Outburst of Comet 17P/Holmes. Astrophysical Journal Letters 677, L63-L66. DOI:  
10.1086/587838
\bibitem[Moreno and Jehin(2025)]{Moreno2025}
Moreno, F., Jehin, E. 2025. Dust shells and dark linear structures on dust tails of historical and recent long-period comets. Astronomy \& Astrophysics, 696, A43. https://doi.org/10.1051/0004-6361/202553986
\bibitem[M\"uller et al.(2024)]{Muller2024}
M\"uller, D. R.. Altwegg, K., Berthelier, J.-J., Combi, M. R., De Keyser, J., et al. 2024. Deciphering cometary outbursts: linking gas composition changes to trigger mechanisms. Monthly Notices of the Royal Astronomical Society 529, 2763-2776. DOI: 10.1093/mnras/stae622
\bibitem[Nissinen et al.(2025)]{Nissinen2025}
Nissinen, M., Gritsevich, M., Weso\l{}owski, M., Ryske, J., and Castro-Tirado, A. J.: Modeling Sublimation Dynamics and Dust Propagation of Comet 17P/Holmes During its 2007 Outburst, EPSC-DPS Joint Meeting 2025, Helsinki, Finland, 7–12 Sep 2025, EPSC-DPS2025-17, https://doi.org/10.5194/epsc-dps2025-17, 2025.
\bibitem[Prialnik et al.(1995)]{Prialnik1995}
Prialnik, D., Brosch, N., Ianovici, D., 1995. Modelling the activity of 2060 Chiron. MNRAS 276, 1148-1154. DOI: 10.1093/mnras/276.4.1148
\bibitem[Reach et al.(2010)]{Reach2010}
Reach, W. T., Vaubaillon, J., Lisse, C. M., Holloway, M., Rho, J., 2010, Icarus, 208, 276-292. DOI: 10.1016/j.icarus.2010.01.020
\bibitem[Ryske et al.(2022)]{Ryske2022}
Ryske, J., Gritsevich, M., Nissinen, M., 2022, 16th Europlanet Science Congress 2022, held 18-23 September 2022 at Palacio de Congresos de Granada, Spain. DOI: 10.5194/epsc2022-60
\bibitem[Sekanina(2007)]{Sekanina2007}
Sekanina, Z., 2007. Comet 17P/Holmes. Central Bureau Electronic Telegrams, No. 1118, 1.
\bibitem[Sekanina(2008)]{Sekanina2008}
Sekanina, Z., 2008. Exploding Comet 17P/Holmes. International Comet Quarterly 30, 3-28. 
\bibitem[Skiff(2018)]{Skiff2018}
Skiff, B., 2018. Outburst of comet 174/Echeclus. J. Br. Astron. Assoc. 128, 51.
\bibitem[Stevenson et al.(2010)]{Stevenson2010}
Stevenson, R., Kleyna, J., Jewitt, D., 2010. Transient Fragments in Outbursting Comet 17P/Holmes. Astronomical Journal 139, 2230-2240. DOI: 10.1088/0004-6256/139/6/2230
\bibitem[Stevenson and Jewitt(2012)]{Stevenson2012}
Stevenson, R., Jewitt, D., 2012. Near-nucleus Photometry of Outbursting Comet 17P/Holmes. Astronomical Journal 144, 138. DOI: 10.1088/0004-6256/144/5/138
\bibitem[Sykes et al.(1986)]{Sykes1986}
Sykes, M. V., Lebofsky, L. A., Hunten, D. M., Low, F., 1986. The Discovery of Dust Trails in the Orbits of Periodic Comets. Science 232, 1115-1117, DOI: 10.1126/science.232.4754.1115
\bibitem[Trigo-Rodr{\'\i}guez et al.(2008)]{Trigo2008}
Trigo-Rodr{\'\i}guez, J. M., Davidsson, B., Montanes-Rodriguez, P., Sanchez, A., Troughton, B., 2008. All-Sky Cameras Detection and Telescope Follow-Up of the 17P/Holmes Outburst. 39th Lunar and Planetary Science Conference, (Lunar and Planetary Science XXXIX), held March 10-14, 2008 in League City, Texas. LPI Contribution No. 1391., p.1627
\bibitem[Trigo-Rodr{\'\i}guez et al.(2008)]{Trigo2008b}
Trigo-Rodr{\'\i}guez J.M., Garc{\'\i}a-Melendo, E., Davidsson, B.J.R., S{\'a}nchez, A., Rodr{\'\i}guez D., Lacruz, J., de los Reyes J.A. and Pastor S.
Outburst activity in comets. I. Continuous monitoring of comet 29P/Schwassmann-Wachmann 1. Astronomy and Astrophysics 485, 599-606.
\bibitem[Trigo-Rodr{\'\i}guez et al.(2010)]{Trigo2010}
Trigo-Rodr{\'\i}guez, J.M., Garc{\'\i}a-Hern{\'a}ndez, D.~A. and S{\'a}nchez, Albert and Lacruz, J. and Davidsson, B. J.R. and Rodr{\'\i}guez D.
Outburst activity in comets - II. A multiband photometric monitoring of comet 29P/Schwassmann-Wachmann 1. MNRAS 409, 1682-1690.
\bibitem[Trigo-Rodr{\'\i}guez et al.(2010)]{Trigo2021}
Trigo-Rodr{\'\i}guez, J. M., Blum, J., 2021. Learning about comets from the study of mass distributions and fluxes of meteoroid streams. Monthly Notices of the Royal Astronomical Society 512, 2277–2289. DOI:  10.1093/mnras/stab2827
\bibitem[Trigo-Rodr{\'\i}guez et al.(2024)]{Trigo2024}
Trigo-Rodr{\'\i}guez, J. M., Sanchez, A., Llenas, J. M., Gritsevich, M., 2024. Photometric Follow-Up of Halley-Type Comet 12P/Pons-Brooks: Outbursts Experienced During the Pre-Perihelion Approach. 55th Lunar and Planetary Science Conference, held 11-15 March, 2024 at The Woodlands, Texas/Virtual. LPI Contribution No. 3040, id.1363
\bibitem[Trigo-Rodr{\'\i}guez et al.(2025)]{Trigo2025}
Trigo-Rodr{\'\i}guez, J. M.,  Souami, D., Gritsevich, M., Weso{\l}owski, M.,  Borisov, G., 2025. Cometary Observations in Light-Polluted Environments: A case study of Interstellar Comet 2I/Borisov.  Astrophysics and Space Science 370, 34. DOI: 10.1007/s10509-025-04424-9
\bibitem[Wesołowski and Gronkowski(2018)]{Wesołowski2018}
Weso\l{}owski, M., Gronkowski, P., 2018. A new method for determining the mass ejected during the cometary outburst - Application to the Jupiter-family comets. New Astronomy 62, p. 55-61. DOI: 10.1016/j.newast.2018.01.006 
\bibitem[Wesołowski(2019)]{Wesołowski2019}
Weso\l{}owski, M., 2019. Impact of light pollution on the visibility of astronomical objects in medium-sized cities in Central Europe on the example of the city of Rzesz$\acute{\mathrm{o}}$w, Poland. Journal of Astrophysics and Astronomy 40, 20. DOI: 10.1007/s12036-019-9586-1
\bibitem[Wesołowski(2020)]{Wesołowski2020}
Weso\l{}owski, M., 2020. Change in the brightness of interstellar comet 2I/Borisov. Planetary and Space Science 194, 105117. DOI:  
10.1016/j.pss.2020.105117
\bibitem[Wesołowski et al.(2020)]{Wesołowski2020}
Weso\l{}owski, M., Gronkowski, P., Tralle, I., 2020. Outbursts of comets at large heliocentric distances: concise review and numerical simulations of brightness jumps. Planetary \& Space Science 184, 104867. DOI: 10.1016/j.pss.2020.104867
\bibitem[Weso\l{}owski(2021)]{Wesołowski2021}
Weso\l{}owski, M., 2021. The influence of the size of ice-dust particles on the amplitude of the change in the brightness of a comet caused by an outburst. Monthly Notices of the Royal Astronomical Society 505, 3525-3536. DOI: 10.1093/mnras/stab1418
\bibitem[Weso\l{}owski(2022a)]{Wesołowski2022a}
Weso\l{}owski, M., 2022a. The rise time of the change of cometary brightness during its outburst. Icarus 375, 114847. DOI: 10.1016/j.icarus.2021.114847
\bibitem[Weso\l{}owski et al.(2022b)]{Wesołowski2022b}
Weso\l{}owski, M., Gronkowski, P., Kossacki, K. J., 2022b. The influence of the porosity of dust particles on the amplitude of the change in the brightness of a comet. Monthly Notices of the Royal Astronomical Society 517, 4950–4958 DOI: 10.1093/mnras/stac2967
\bibitem[Wesołowski(2023)]{Wesołowski2023}
Weso\l{}owski, M., 2023. The increase in the surface brightness of the night sky and its importance in visual astronomical observations. Scientific Reports 13, 17091. DOI: 10.1038/s41598-023-44423-w
\bibitem[Wesołowski and Potera(2024)]{Wesołowski2024}
Weso\l{}owski, M., Potera, P., 2024. Determination of bolometric albedo based on spectroscopic measurements for selected dust analogues and its impact on the change of cometary brightness during its outburst. Astronomy \& Astrophysics  686, A248. DOI: 10.1051/0004-6361/202449573
\bibitem[Wesołowski (2025)]{Wesołowski2025}
Weso\l{}owski, M., 2025. The flux-nuclear mechanism as the cause of cometary outbursts – a solution to an old problem. Monthly Notices of the Royal Astronomical Society 539, 939-948. DOI: 10.1093/mnras/staf551
\bibitem[Wesołowski et al.(2025)]{Wesołowski2025a}
Weso\l{}owski M., Carson P., Gritsevich M., 2025. Surface Correction Values for the Outbursts of Comet 29P/Schwassmann–Wachmann. Publications of the Astronomical Society of the Pacific,  137, 114403. DOI: 10.1088/1538-3873/ae170c
\end{thebibliography}
\end{document}